\documentclass[longauth, letter]{aa}  

\graphicspath{{./}{figures/}}

\usepackage{multirow}
\usepackage{txfonts}
\usepackage{graphicx}
\usepackage[outdir=./]{epstopdf}
\usepackage{natbib}
\usepackage{xcolor, color}
\usepackage{booktabs}
\usepackage{threeparttable}
\usepackage{hyperref}
\hypersetup{breaklinks=true, linkcolor=red, citecolor=blue, filecolor=cyan, urlcolor=magenta}
\usepackage{xurl}
\usepackage{comment}

\DeclareUnicodeCharacter{2212}{-}
\newcommand{\kms}{km s$^{-1}$}

\newcommand{\meth}{CH$_3$OH}

\begin{document}

%\title{FAUST ??. The inner 300 au of the IRAS 4A2 protostar mapped in CH$_3$OH and SiO: the hot corino or the base of a wind?}
\title{FAUST XV. A disk wind mapped by CH$_3$OH and SiO in the inner 300 au of the NGC 1333 IRAS 4A2 protostar.}

%\correspondingauthor{Marta De Simone}
%\email{marta.desimone@eso.org}
% 0000-0001-5659-0140 orcid marta
\author{
M. De Simone\inst{1,2}
\and L. Podio \inst{2}
\and L. Chahine \inst{3}
\and C. Codella \inst{2,3}
\and C. J. Chandler \inst{4}
\and C. Ceccarelli  \inst{3}
\and A. L\'{o}pez-Sepulcre\inst{3,5}
\and L. Loinard\inst{6,7}
\and B. Svoboda \inst{4}
\and N. Sakai \inst{8}
\and D. Johnstone\inst{9,10}
\and F. M\'enard \inst{3}
\and Y. Aikawa\inst{11}
\and M. Bouvier\inst{12}
\and G. Sabatini\inst{2}
\and A. Miotello\inst{1}
\and C. Vastel\inst{13}
\and N. Cuello\inst{3}
\and E. Bianchi \inst{14}
\and P. Caselli\inst{15}
\and E. Caux\inst{13}
%\and Aurore Dur\'{a}n\inst{6}
\and T. Hanawa \inst{16}
\and E. Herbst\inst{17}
\and D. Segura-Cox\inst{15,18}
\and Z. Zhang \inst{8}
\and S. Yamamoto \inst{19}
%\and the FAUST Team
}

\institute{
ESO, Karl Schwarzchild Srt. 2, 85478 Garching bei München, Germany \\ \email{marta.desimone@eso.org}
\and 
INAF, Osservatorio Astrofisico di Arcetri, Largo E. Fermi 5, 50125 Firenze, Italy
\and
Univ. Grenoble Alpes, CNRS, IPAG, 38000 Grenoble, France
\and 
National Radio Astronomy Observatory\footnote{\scriptsize The National Radio Astronomy Observatory is a facility of the National Science Foundation operated under cooperative agreement by Associated Universities, Inc.}, 1003 Lopezville Rd, Socorro, NM 87801, USA
\and
Institut de Radioastronomie Millim\'etrique (IRAM), 300 rue de la Piscine, 38406 Saint-Martin-d'H\`eres, France
\and
Instituto de Radioastronom\'{i}a y Astrof\'{i}sica, Universidad Nacional Aut\'{o}noma de M\'{e}xico, A.P. 3-72 (Xangari), 8701, Morelia, Mexico 
\and
Instituto de Astronom\'{i}a, Universidad Nacional Aut\'{o}noma de M\'{e}xico, Ciudad Universitaria, A.P. 70-264, Cuidad de M\'{e}xico 04510, Mexico 
\and
The Institute of Physical and Chemical Research (RIKEN), 2-1, Hirosawa, Wako-shi, Saitama 351-0198, Japan
\and
NRC Herzberg Astronomy and Astrophysics, 5071 West Saanich Road, Victoria, BC, V9E 2E7, Canada
\and 
Department of Physics and Astronomy, University of Victoria, Victoria, BC, V8P 5C2, Canada
\and 
Department of Astronomy, The University of Tokyo, 7-3-1 Hongo, Bunkyo-ku, Tokyo 113-0033, Japan 
\and 
Leiden Observatory, Leiden University, P.O. Box 9513, 2300 RA Leiden, The Netherlands
\and
IRAP, Universit\`{e} de Toulouse, CNRS, CNES, UPS, Toulouse, France 
\and 
Excellence Cluster ORIGINS, Boltzmannstraße 2, 85748, Garching bei Mu\"unchen, -- Germany
\and
Center for Astrochemical Studies, Max-Planck-Institut f\"{u}r extraterrestrische Physik (MPE), Gie{\ss}enbachstr. 1, D-85741 Garching, Germany
\and 
Centre for Frontier Science, Chiba University, 1-33 Yayoi-cho, Inage-ku, Chiba 263-8522, Japan
\and
Department of Chemistry, University of Virginia, McCormick Road, PO Box 400319, Charlottesville, VA 22904, USA
\and 
The University of Texas at Austin, 2515 Speedway, Austin, Texas 78712, USA
\and
The Graduate University for Advanced Studies (SOKENDAI), Shonan Village, Hayama,
Kanagawa 240-0193, Japan
}
%\author[0000-0003-1514-3074]{C. Codella}
%\author[0000-0002-7570-5596]{Claire J. Chandler}
%\author[0000-0001-9664-6292]{Cecilia Ceccarelli}
%\author[0000-0002-3297-4497]{Nami Sakai}
%\author[0000-0002-9865-0970]{Satoshi Yamamoto}

\titlerunning{FAUST XVI.  A wide disk wind mapped by CH$_3$OH and SiO in the inner 300 au of IRAS 4A2.}
\authorrunning{De Simone et al.}

\abstract
% context heading (optional)
{Understanding the connection between outflows, winds, accretion and disks in the inner protostellar regions is crucial for comprehending star and planet formation process. }
  % aims heading (mandatory)
{We aim to we explore the inner 300 au of the protostar IRAS 4A2 as part of the ALMA FAUST Large Program.}
 % methods heading (mandatory)
{We analysed the kinematical structures of SiO and \meth \ emission with 50 au resolution.} % in the inner 300 au of the NGC 1333 IRAS 4A2 protostar.}
 % results heading (mandatory)
{The emission arises from three zones: 
i) a very compact and unresolved region ($<$50 au) dominated by the ice sublimation zone, at $\pm$1.5 \kms \ with respect to v$_{\rm sys}$, traced by methanol;
ii) an intermediate region (between 50 au and 150 au) traced by both SiO and CH$_3$OH, between 2 and 6 \kms \ with respect to v$_{\rm sys}$, with an inverted velocity gradient (with respect to the large scale emission), whose origin is not clear;
iii) an extended region ($>$ 150 au) traced by SiO, above 7 \kms with respect to v$_{\rm sys}$, and dominated by the outflow. 
In the intermediate region we estimated a \meth/SiO  abundance ratio of about 120--400 and a SiO/H$_2$ abundance of 10$^{-8}$.
We explored various possibilities to explain the origin of this region such as, rotating disk/inner envelope, jet on the plane of the sky/precessing, wide angle disk wind.  
}
% conclusions heading (optional)
{%This intermediate region with an inverse velocity gradient could trace the base of a disk wind that crosses the plane of the sky and sputter the grains releasing SiO and CH$_3$OH. 
We propose that \meth \ and SiO in the inner 100 au probe the base of a wide-angle disk wind. The material accelerated in the wind crosses the plane of the sky, giving rise to the observed inverted velocity gradient, and sputtering the grain mantles and cores releasing CH$_3$OH and SiO.
This is the first detection of a disk wind candidate in SiO, and the second ever in \meth.

% {add something to strengthen the letter, something like For the first time, we detect the disk wind in the IRAS 4A2 protostar, a young class 0 source, in SiO and \meth \. New disk wind tracers.}
%{maybe not needed here in the abstract} We emphasise that higher angular resolution observations (down to 10 au) will be essential to reveal the disk (unresolved at 50 au resolution) and complete the overall picture of the system.
}

% the AAS Journals, the Astrophysical Journal (ApJ), the Astrophysical Journal Letters (ApJL), and Astronomical Journal (AJ), all have a 250 word limit for the abstract


\maketitle
%
%-------------------------------------------------------------------

\section{Introduction} \label{sec:intro}
%In the classical star formation scenario, the stage after the collapse of a molecular cloud is characterized by 
Solar-like stars form from an accreting object deeply embedded in a dense envelope that drives bipolar jets. To allow the accretion from the disk onto the protostar, the angular momentum is extracted from the disk by jets/outflows or disk winds   
%In the early stages of their formation, Solar-like stars are accreting objects deeply embedded in a dense envelope. The forming star drives bipolar jets, which are removing the angular momentum allowing the accretion from the disk onto the protostar 
\citep[e.g.,][]{shu_star_1987}. Understanding the accretion and ejection mechanisms and their impact on the physical and chemical structure of protostars is of paramount importance to our comprehension of the evolution of the forming stellar/planetary system. 

Low-mass Class 0 protostars \citep[$\sim$10$^4$ yr;][]{andre_prestellar_2000} are the most studied objects to investigate both the accreting region and the molecular outflows. 
Some protostars hosts hot ($>$100 K), compact ($<$100 au), dense ($>10^7$ cm$^{-3}$) regions rich in iCOMS \citep[interstellar Complex Organic Molecules; ][]{herbst_complex_2009,ceccarelli_seeds_2017}, where the ice mantle sublimation dominates the chemistry \citep[called hot corinos; ][]{ceccarelli_hot_2004}. 
However, not every protostar possesses a hot corino \citep[e.g.,][]{de_simone_glycolaldehyde_2017, belloche_questioning_2020, bouvier_chemical_2022, yang_constraining_2020, de_simone_hot_2020, de_simone_tracking_2022}, and their physical and chemical structure is still not clear. 
%{MdS: I think I remove the following paragraph... it is not important for the scope of the letter, and we have to shrink and go to the point.}
%However, only about 25 iCOMs-rich hot corinos are known \citep[e.g.,][]{de_simone_glycolaldehyde_2017, belloche_questioning_2020, bouvier_chemical_2022, yang_perseus_2021}, and several scenario have been proposed to explain their scarcity: obscuration of the molecular emission due to the optically thick dust \citep{de_simone_hot_2020}, lower gas temperatures due to shadowing by small scale structures \citep{van_gelder_importance_2022, nazari_importance_2022, nazari_importance_2023}, or different grain mantle composition  \citep{sakai_warm_2013,de_simone_tracking_2022}. Beyond that, the physical and chemical structure of a hot corino is still not known. %Indeed, they are very compact (below 100 au), and require high angular resolution observation at high sensitivity. 
Among the few known hot corinos, only a handful are spatially resolved \citep[SVS13-A, HH212, IRAS 16293A, B335;][]{ bianchi_streamers_2023, Lee_stratified_2022, maureira_dust_2022, okoda_chemical_2022}. 
Here, the iCOMs emission seems to probe different layers of the hot corino or selected regions associated to accretion shocks and/or hot spots. 
%In these sources the molecular emission seems to suggest a stratified structure or to follow accretion shocks/hot spots.  

On the other hand,  jets and outflows driven by protostellar sources have been extensively studied, especially at scales larger than 500 au  \citep[e.g.,][]{ray_toward_2007, arce_molecular_2007, franck_jets_2014}. 
However, their origin is still debated. 
%Models predict that they are launched by a magneto centrifugal mechanisms which extracts the material from a large range of disk radii: from close to the dust truncation radius \citep[X-wind models:][]{shu_magnetocentrifugally_1994, arce_molecular_2007, shang_jets_2007} out to $\sim10-100$ au  \citep[disk-wind models:][]{blandford_hydromagnetic_1982, konigl_disk_2000, tsukamoto_role_2022}. 
 {Models predict that magneto centrifugal mechanisms extracts material from the disk from two regions: close to the dust truncation radius by a stellar magnetosphere \citep[X-wind models:][]{shu_magnetocentrifugally_1994, arce_molecular_2007, shang_jets_2007}, and a wider range of radii outside this dust truncation radius \citep[disk-wind models:][]{blandford_hydromagnetic_1982, konigl_disk_2000, tsukamoto_role_2022}. }

%Observations hint that they may originate from a larger disk region \citep[e.g.,][]{tabone_alma_2017,zhang_rotation_2018,ai_unified_2024}.
%In  both scenarios the wind from the inner disk is accelerated and collimated into a high-velocity jet which shocks the surrounding material. A slower, wide-angle outflow may originate from the outer disk region and/or due to entreinment of cloud material along the jet.
These winds have been considered as a solution to the angular momentum problem in the inner 50 au of the disks \citep[e.g.,][]{pascucci_role_2023}.  {However, only a handful of resolved disk winds observations are available \citep[e.g.,][]{tabone_alma_2017, zhang_rotation_2018, louvet_hh30_2018, de_valon_alma_2020, lee_first_2021,lopez-vazquez_multiple_2024}. }These type of observations provide crucial information (e.g., geometry, kinematical and chemical structure) to constrain disk wind models, and to retrieve their origin, and the interplay between accretion and ejection mechanisms. 

In this Letter we explore the inner 300 au of the IRAS 4A2 protostar as part of the ALMA FAUST\footnote{Fifty AU STudy of the chemistry in the disk/envelope systems of Solar-like protostars; \url{http://faust-alma.riken.jp}.} \citep[][]{codella_enlightening_2021} Large Program in SiO and CH$_3$OH. We study the kinematics of the gas in the transition zone between a possible disk and the ouflowing material, and present, for the first time, a wide disk wind candidate for NGC 1333 IRAS 4A2 traced by SiO.
%IRAS 4A2 is a known Class 0 hot corino launching a bipolar molecular outflow, and is therefore the ideal target to investigate the kinematics and the interaction between the outflow, the hot corino, and the disk/inner envelope, at small scales.

%\subsection{The NGC 1333 IRAS 4A protostellar system}
\paragraph{\textit{The target:}} NGC 1333 IRAS 4A is a very well studied system in Perseus \citep[$\sim$300 pc away;][]{zucker_mapping_2018}. It is composed of two protostars: IRAS 4A1 (4A1 hereinafter), the brightest in mm continuum, and IRAS 4A2 (4A2 hereinafter), separated by about $1\farcs8$ \citep[about 540 au;][]{looney_unveiling_2000,santangelo_jet_2015,lopez-sepulcre_complex_2017,tobin_vlaalma_2018}. 
They are surrounded by a common envelope of about 8 M$_\odot$ \citep{maury_characterizing_2019} and have a total bolometric luminosity of about 9 L$_\odot$ \citep{kristensen_water_2012,karska_water_2013}. Note that the system is also known to show, at sub-mm wavelengths on years timescales, a $\sim$10\% peak to peak variability that can slightly affect the luminosity \citep{lee_jcmt_2021, mairs_jcmt_2024}.
Their mm dust emission is optically thick in the central region \citep[][]{maury_characterizing_2019, galametz_low_2019, li_systematic_2017, chia-lin_resolving_2020, guerra-alvarado_iras4a1_2023} to completely obscure (for 4A1) or partially absorb (for 4A2) the mm iCOMs emission \citep{de_simone_hot_2020}. 
% {MdS: shorten this, or remove it!}Indeed, while 4A2 has always been a well known hot corino \citep{bottinelli_complex_2004, taquet_constraining_2015, de_simone_glycolaldehyde_2017, lopez-sepulcre_complex_2017, quitian-lara_decoding_2024}, 4A1 was not. However, in \citeyear{de_simone_hot_2020}, \citeauthor{de_simone_hot_2020} have shown the presence of methanol at cm wavelengths towards both sources with similar intensities, assessing that the dust was so optically thick at mm wavelengths to obscure totally, for 4A1, and partially, for 4A2, the molecular line emission.
The two protostars emit large bipolar jet/outflows, deeply studied with several tracers, such as CO, SiO, SO, HCN and some iCOMs \citep[including methanol; e.g.,][]{lefloch_widespread_1998, choi_radio_2011, ching_helical_2016, de_simone_seeds_2020, chuang_alma_2021}. 
The southern lobes are blue-shifted, and the northern ones red-shifted and they cover a large velocity range \citep[up to 60 \kms;][]{choi_variability_2005,santangelo_jet_2015,taquet_seeds_2020}.
%, with a peculiar bending toward the northeast at about $20''$ from the protostars,
%With VLA, \citet{choi_variability_2005}  mapped the high-velocity component at large scale (about 1$'$) using SiO at 2$''$ angular resolution. With IRAM/PdBI observation at $\sim1''$ angular resolution using CO, SiO, and SO, a smaller scale ($<20''$) was mapped by \citet{santangelo_jet_2015} tracing different velocity components (up to 60 \kms). With IRAM/NOEMA SOLIS observations at at $\sim 0\farcs8-2''$ angular resolution, \citet{taquet_seeds_2020} mapped the 30$''$ scales with OCS, SO, SO$_2$ and CS, disentangling the northern red-shifted outflows. 
However, none of the previous studies resolve the inner 200 au (below $0\farcs7$). 
The FAUST data provide the superb spatial resolution of 50 au, that allows us to explore for the first time these inner regions using selected chemical tracers. 

\begin{figure*}
    \centering
    \includegraphics[width=0.9\textwidth]{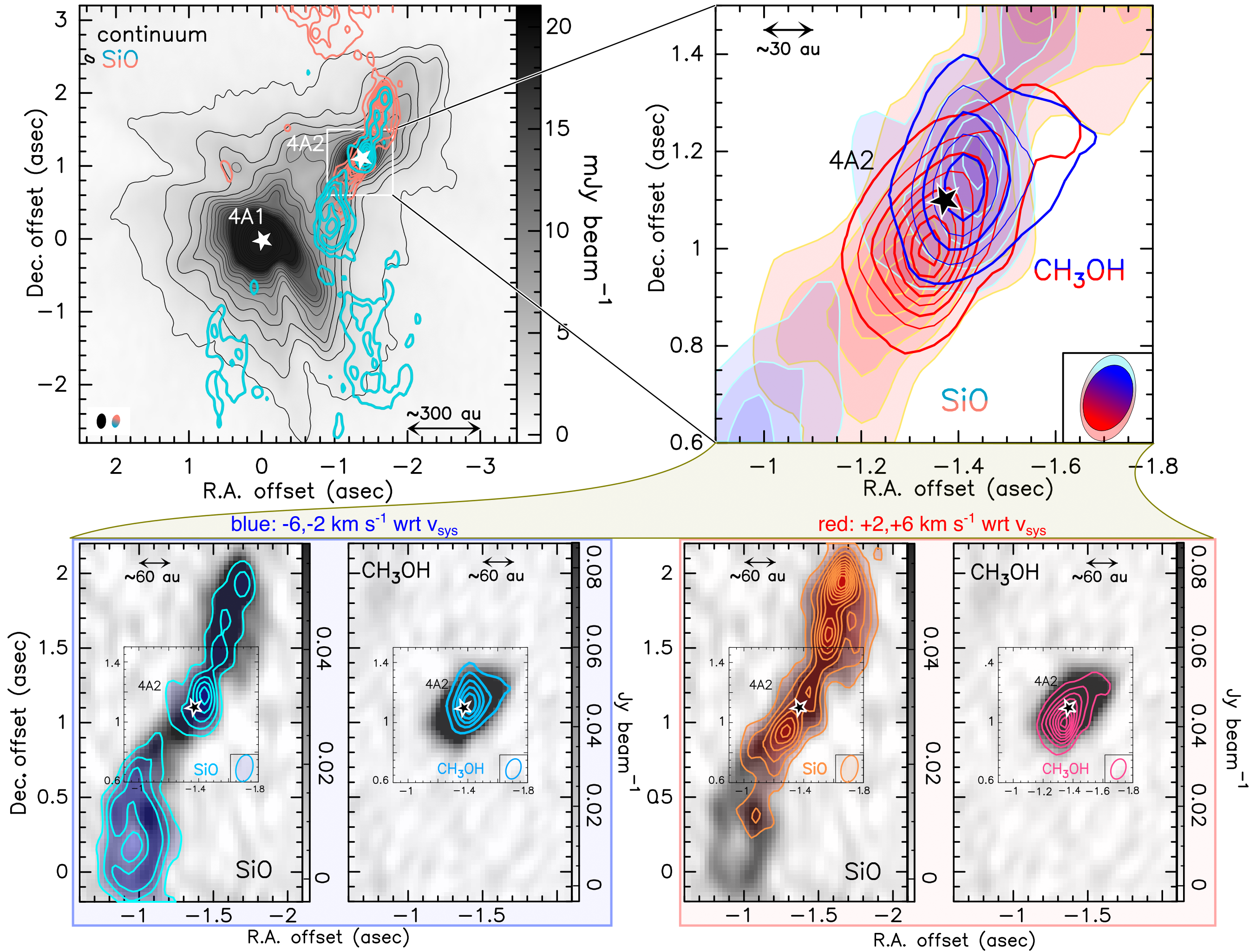}
    \caption{\textit{Upper left:} Superposition of the 1.3 mm continuum (grey scale, up to 20\% of maximum flux, with contours starting at 15$\sigma_c$ with step of 20$\sigma_c$, where $\sigma_c=8\times10^{-2}$ mJy beam$^{-1}$) with SiO red- and blue-shifted emission (salmon and cyan contours, respectively, from 3$\sigma$ with steps of 4$\sigma$, where $\sigma=2$ mJy beam$^{-1}$ \kms). %Beams are reported in the lower left. %(continuum: $0\farcs21\times0\farcs13$ ($-5^\circ$), SiO: $0\farcs18\times0\farcs11$ ($-13^\circ$)).
    \textit{Upper right:} Zoom-in on IRAS 4A2. Superposition of the red- and blue-shifted emission of SiO (salmon and cyan color shaded contours) and CH$_3$OH (red and blue contours). Contours starts at 3$\sigma$ with steps of 4$\sigma$ for SiO and 5$\sigma$ for CH$_3$OH ($\sigma=2$ mJy beam$^{-1}$ \kms).
    Beams are shown in the lower corners. %(SiO: $0\farcs18\times0\farcs11$ ($-13^\circ$), CH$_3$OH: $0\farcs14\times0\farcs096$ ($-21^\circ$)).
    \textit{Lower panels:} Blue- and red-shifted emission for SiO and CH$_3$OH shown separately. The beams are shown in the inner corner. 
    In all panels, white (and black) stars mark the protostars (IRAS 4A1 and IRAS 4A2). The red- and blue-shifted emissions are integrated between $+2$ and $+$6 km s$^{-1}$, and between $-6$ and $-2$ \kms \ with respect to the systemic velocity \citep[$\rm v_{sys}=+6.8$ km s$^{-1}$;][]{choi_high-resolution_2001}, respectively.
     {The moment 0 between -6 and 6 \kms \  for each species is shown in greyscale.}}
    \label{fig:cont+sio+ch3oh}
    \vspace{-0.5cm}
\end{figure*}

\begin{figure}[ht]
    \centering
    \includegraphics[width=0.7\columnwidth]{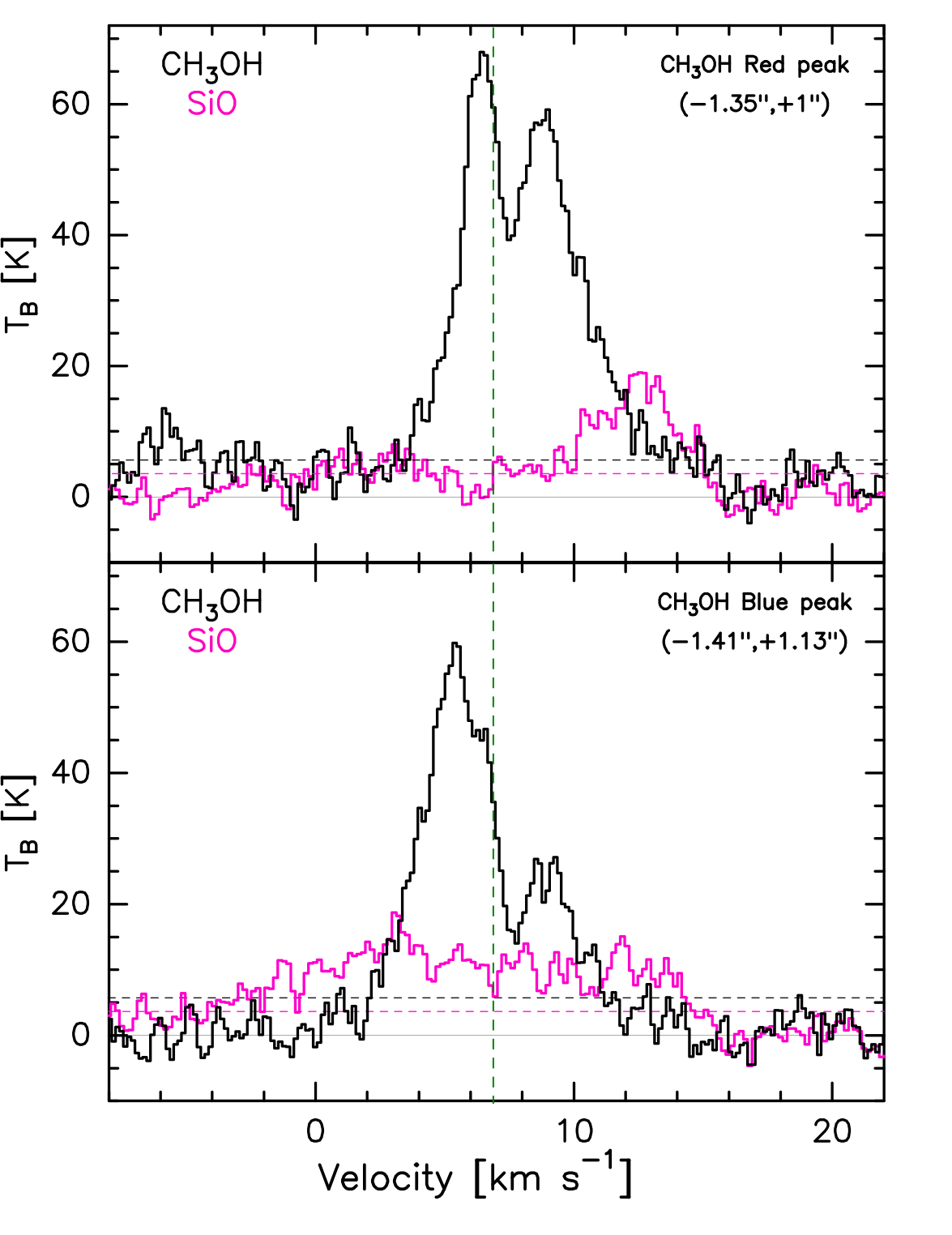}
    \caption{SiO (magenta) and CH$_3$OH (black) spectra extracted at the blue- and red-shifted peaks of CH$_3$OH %and SiO 
    (offset with respect to the map phase center are in the upper right corner).
    The green dashed vertical line marks the v$_{sys}$ \citep[$\sim$6.8 \kms][]{choi_high-resolution_2001}. The 3$\sigma$ level is reported in black (\meth) and magenta (SiO) dashed horizontal lines.  
    }
    \label{fig:sio&ch3oh_spectra}
    \vspace{-0.5cm}
\end{figure}

\begin{table*}
    \centering
    \begin{threeparttable}
    \setlength{\tabcolsep}{3.5pt}
    \renewcommand{\arraystretch}{1.3}
    \caption{\textit{Top:} Spectral properties of \meth \ and SiO by \citet{xu_torsionrotation_2008} and  \citet{muller_rotational_2013}, respectively, from the CDMS database \citep{muller_cologne_2005}.
    \textit{Bottom:} Integrated Area in the blue (from 0.8 to 4.8 \kms) and red (from 8.8 to 12.8 \kms) velocity ranges from the spectra extracted at the blue- and red-shifted peaks of CH$_3$OH and SiO (see Figure \ref{fig:cont+sio+ch3oh} and \ref{fig:sio&ch3oh_spectra}) and corresponding column densities.
    }    \label{tab:spectral properties}
    \begin{tabular}{c|cccccccc}
        \hline
        \hline
        Line & \multicolumn{2}{c}{Transition} & {Frequency} & E$\rm_u$ & {$\rm \log(A_{ul}/s^{-1})$} & g$\rm_u$ & \multicolumn{2}{c}{Beam} \\ 
        & \multicolumn{2}{c}{}& {[GHz]} & {[K]} & & & \multicolumn{2}{c}{$"\times"(^\circ)$} \\
        \hline
         SiO & \multicolumn{2}{c}{5-4} & {217.10498} & {31} & {-3.28} & 11 & \multicolumn{2}{c}{$0.18\times0.11 (-13)$}   \\
         CH$_3$OH & \multicolumn{2}{c}{5$_{1,4}-4_{1,3}$ A} & {243.91579} & {50} & {-4.22} & 44 & \multicolumn{2}{c}{$0.14\times0.096 (-21)$}\\
        \hline
        \hline
         & \multicolumn{8}{c}{Velocity-integrated intensity ($\int I\Delta$v)$^a$ \ \& \ \ Column Density (N$_{\rm col}$)$^b$} \\
        \hline
          Position & \multicolumn{2}{c}{CH$_3$OH Red Peak} & \multicolumn{2}{c}{SiO Red Peak} & 
         \multicolumn{2}{c}{CH$_3$OH Blue Peak} & 
         \multicolumn{2}{c}{SiO Blue Peak}  \\
         Offset$^c$ & \multicolumn{2}{c}{$(-1\farcs35, +1")$} &
         \multicolumn{2}{c}{$(-1\farcs27, +0\farcs93)$} &
         \multicolumn{2}{c}{$(-1\farcs41, +1\farcs13)$} & 
         \multicolumn{2}{c}{$(-1\farcs44, +1\farcs19)$} \\
         \hline
          & $\int I\Delta$v & N$_{\rm col}$ & $\int I\Delta$v & N$_{\rm col}$ & $\int I\Delta$v & N$_{\rm col}$ & $\int I\Delta$v & N$_{\rm col}$ \\ 
          & [K \ km s$^{-1}$] & [cm$^{-2}$] &  [K \ km s$^{-1}$] & [cm$^{-2}$] &  [K \ km s$^{-1}$] &[cm$^{-2}$] &  [K \ km s$^{-1}$] & [cm$^{-2}$] \\ \hline
         SiO & 44(6) & 8.7(1.1)$\times 10^{13}$ & 54(8) & 10(2)$\times 10^{13}$ & 
         51(8) & 10(2)$\times 10^{13}$ & 43(6) & 8.5(1.1)$\times 10^{13}$   \\
         CH$_3$OH &  118(17) & 37(5)$\times 10^{15}$ & 37(6) & 12(2)$\times 10^{15}$ & 67(10) & 21(3)$\times 10^{15}$ & 52(8) & 16(2)$\times 10^{15}$ \\
         %$\rm \frac{SiO}{CH_3OH}$ & 
         \hline
         & $\int I\Delta$v & N$_{\rm col}$ & $\int I\Delta$v & N$_{\rm col}$ & $\int I\Delta$v & N$_{\rm col}$ & $\int I\Delta$v & N$_{\rm col}$ \\ 
         \hline
         CH$_3$OH/SiO &  2.7(0.8) & 420(110) & 0.7(0.2) & 120(40) & 1.3(0.4) & 210(70) & 1.2(0.3) & 190(50)
         \\
         \hline 
    \end{tabular}
    \begin{tablenotes}
         \item $^a$ Errors on the integrated intensity include 15\% of calibration error and spectral baseline determination.
         %\item $^b$ The column density has been computed assuming an excitation temperature of 100 K. Note that a difference in temperature of 50 K translates into about a factor 2 different in column density. 
         \item $^b$ N$_{\rm col}$ has been computed assuming an $T_{ex}$ of 100 K. A difference in $T_{ex}$ of 50 K translates into about a factor 2 in N$_{\rm col}$. 
         \item $^c$ with respect to the map phase center. 
    \end{tablenotes}
  \end{threeparttable}
\end{table*}
\section{Observations} \label{sec:obs}
The observation of NGC 1333 IRAS 4A we present here are part of the Cycle 6 ALMA Large Program FAUST (PI. S. Yamamoto, 2018.1.01205.L). 
They were performed between October 2018 and September 2019 with baselines for the 12-m array between 15.1 m and 3.6 km.
The bandpass and flux calibrator was J0237+2848, while J0336+3218 and J0328+3139 were used for the phase calibration. 
The map phase center is at R.A. (J2000) = 03h29m10s.539,
Dec. (J2000)= 31$^\circ$13$'30\farcs92$. 
%More details on the calibration can be found in Chahine et al. (submitted). %, here we summarize the most important details.
For this work we used CH$_3$OH (5$_{1,4}-4_{1,3}$A) and SiO ($5-4$) lines at 243.916 GHz and 217.105 GHz, respectively, in the high resolution spectral windows { ($\Delta\nu=0.123$ kHz, corresponding to about 0.15 \kms \ and  0.17 \kms, for CH$_3$OH and SiO, respectively)}. 
The data were calibrated using the ALMA calibration pipeline within CASA\footnote{\url{https://casa.nrao.edu}} with an additional calibration routine to correct for the Tsys normalization issue\footnote{%\href{https://help.almascience.org/kb/articles/what-are-the-amplitude-calibration-issues-caused-by-alma-s-normalization-strategy}{almascience\_articles/normalization-strategy}
\url{https://help.almascience.org/kb/articles/what-are-the-amplitude-calibration-issues-caused-by-alma-s-normalization-strategy}
}. After aligning data from different setups, phase-only self-calibration was performed on the continuum, generated using a careful manual detection of line-free channels. The continuum model was then subtracted from the visibilities prior to imaging the line data and a continuum image was then created in CASA.
%{\color{magenta} {To be checked by Claire:} The spectral window containing the SiO line lacks of line-free channels. Therefore, the continuum has been subtracted using the wide spectral window one and correcting for the spectral index. }
%The one shown in Figure \ref{fig:cont+sio+ch3oh} is the one at 1.3mm using a 0.3 robust parameter resulting in a synthesized beam of $0\farcs21\times0\farcs13$ (-5$^\circ$). 
The resulting continuum-subtracted line cube was created %combining the two available configurations (TM1 and TM2) and 
combining the two available ALMA-12m configurations (i.e. C43-3, C43-6), to have the best angular resolution, and imaged with the IRAM-GILDAS\footnote{\url{https://www.iram.fr/IRAMFR/GILDAS/}} package, using a 0.56 robust parameter \footnote{https://www.iram.fr/IRAMFR/GILDAS/doc/pdf/map.pdf} with a resulting synthesized beam of $0\farcs18\times0\farcs11$ (-13$^\circ$) and $0\farcs14\times0\farcs096$ (-21$^\circ$) for the SiO and CH$_3$OH spectral windows, respectively. The cubes were then primary beam corrected. 
We estimate the absolute flux error of 15\% that includes the calibration uncertainty and an additional error for the spectra baseline determination.

\section{The kinematical structure of the IRAS 4A inner regions }

%The large scale outflow traced by SiO and the velocity ranges covered by the FAUST data are shown in Figure \ref{fig:sio_largescale} for completeness. The study of the large scale structure is out of the scope of this letter and presented in Chahine et al. (submitted).

Figure \ref{fig:cont+sio+ch3oh} shows the dust continuum emission at 1.3 mm of NGC 1333 IRAS 4A overlaid with the SiO 5-4 low velocity emission \citep[in the ranges $(-6,-2)$ and $(+2,+6)$  \kms \ with respect to the v$_{\rm sys}$; $\rm v_{sys}=6.8$ km s$^{-1}$;][]{choi_high-resolution_2001}, with a zoom-in on SiO and CH$_3$OH emission lines (listed in Table 1) at $<1''$ ($<$300 au) scale around IRAS 4A2. %The first findings that pops up is that both methanol and SiO clearly show an opposite velocity gradient at $<0.5''$ with respect to the large scale one. 
At large scale (above 1$''$ from IRAS 4A2, Figure \ref{fig:sio_largescale}), both SiO and \meth \ trace blue-shifted emission in the south (up to -33 \kms \ with respect to v$_{\rm sys}$), and red-shifted emission in the north (up to +35 \kms \ with respect to v$_{\rm sys}$). 
However, at small scales the kinematics is different: below 300 au, both SiO and \meth \ emit in the velocity ranges ($-6, -2)$ \kms \ and  ($+2, +6)$ \kms \ with respect to v$_{\rm sys}$ (see also channel maps in Figure \ref{fig:sio_chanmaps}, and \ref{fig:ch3oh-chanmaps}) with a velocity gradient inverted with respect to the large scale.
Specifically, CH$_3$OH shows blue-shifted emission towards the northeast and red-shifted emission towards the southwest, that peaks at $\sim0\farcs15$ from IRAS 4A2 and extends out to $\sim 0\farcs3$ .
The SiO emission has an S-shape structure that follows the continuum emission from the dust grains in the circumstellar envelope. 
In the inner 300 au, it shows both blue- and red-shifted emission on both sides. 
However, in the inner $0\farcs3$ ($\sim$90 au) the SiO distribution resembles that of methanol, with an inverted velocity gradient with respect to the large scale emission. 
At larger distances, i.e. from $0\farcs3$ to $1''$, the southern lobe is only blue-shifted (in agreement with the large scale), while the northern lobe is mainly red-shifted with a tiny blue-shifted contamination (that dominates closer to the v$_{sys}$; at about $-2$ \kms, Figure \ref{fig:sio_chanmaps}). 
 {We also performed PV diagrams along the methanol and SiO PA and perpendicular to the emission (see Figure \ref{fig:regions_pv},\ref{fig:pvs_sio+ch3oh}, \ref{fig:sio_pv_diagrams}). The inversion of velocity is clearly shown, but the PV diagrams are limited in spatial/angular resolution to reveal any other kinematical structure.  
}
%REFRASED At a first sight, it seems to show both blue- and red-shifted emission. However, at $0\farcs5$ away from the protostar there is a component that behave like methanol, i.e. with an inverted velocity gradient. From $0\farcs5$ to $1''$ away from the protostar, the southern lobe is only blue-shifted (in agreement with the large scale), while the northern lobe is mainly red-shifted with a tiny blue-shifted contamination (that dominates closer to the systemic velocity; $-2,-3$ \kms with respect to the v$_{\rm sys$). 

%We highlight that both CH$_3$OH and SiO have an inverted velocity gradient at $< 0\farcs3$ with respect to the large scale. % but a different spatial distribution. Indeed, 
%{mds: move later...}Additionally, the moment 0 maps in Figure \ref{fig:cont+sio+ch3oh} show that methanol peaks closer to the source than SiO in both blue and red ranges (by $0.1''$, less than a beam). %, but the shift between the two species is detected consistently in the maps integrated on the same velocity intervals). 
%in order to understand if this difference is significative, given it is not resolved out, %and if it can help us understanding the origin of this emission, 
%To understand if this difference is significative, we estimated the 
 {Figure \ref{fig:sio&ch3oh_spectra} shows the spectra extracted at the position of the \meth\, red- and blue-shifted emission peaks (coordinate offsets reported in Table \ref{tab:spectral properties}). 
The two peaks were identified as the pixel with the highest flux level in the integrated maps of \meth \ in the $(-6,-2)$ \kms \ and $(+2,+6)$ \kms \  ranges (see Figure \ref{fig:cont+sio+ch3oh}). }
The CH$_3$OH spectrum peaks at low velocity ($\sim\pm1.5$ \kms \ with respect to v$_{\rm sys}$) and is likely dominated by the ice mantle sublimation region (i.e., the hot corino). The SiO spectrum, instead, peaks at higher velocity (about $\pm7$ \kms \ with respect to v$_{\rm sys}$) and it is likely dominated by jet/outflow emission due to grain sputtering/shattering. 
In the $(-6,-2)$ \kms \ and $(+2,+6)$ \kms \  ranges the spectra of CH$_3$OH and SiO overlap. In these ranges we observe the inverted velocity gradient.
%The inverted velocity gradient observed in the integrated maps covers the $(-6,-2)$ \kms \ and $(+2,+6)$ \kms \  ranges. Here, the spectra of the two species overlap.  
We measured the intensity ratio CH$_3$OH/SiO (Table \ref{tab:spectral properties}), integrated on the $(-6,-2)$ \kms \ and $(+2,+6)$ \kms \  ranges, and the corresponding abundance ratio, estimated assuming LTE (Local Thermodynamic Equilibrium) optically thin emission and an excitation temperature of 100 K (typical for both hot corino and shocked gas), in the red- and blue- shifted emission peaks of both SiO and \meth. 
 {The emission peaks of SiO were identified, as for \meth, as the pixel with the highest flux level in the integrated maps (Figure \ref{fig:cont+sio+ch3oh}).}
We found that, in general, CH$_3$OH is more abundant than SiO by about two orders of magnitude. 
Additionally, the moment 0 maps in Figure \ref{fig:cont+sio+ch3oh} show that methanol peaks closer to the source than SiO in both blue and red ranges (by $0.1''$, less than a beam). Even if unresolved, this spatial shift in the red-shifted component seems to be significant. This is supported by the fact that the \meth/SiO integrated intensity and abundance ratio estimated by extracting the spectra at the SiO peak are both about a factor 4 smaller than that estimated at the \meth \ peak (Table \ref{tab:spectral properties}).
%and that in the southern red-shifted component the shift between the SiO and \meth \ emission peaks is significant. 
%This is supported by the fact that the \meth/SiO  intensity and abundance ratio estimated by extracting the spectra at the SiO peak are about a factor 4 smaller than the one estimated at the \meth \ peak.
%, resulting in a difference in \meth/SiO of about a factor 4. 
Finally, in IRAS 4A2 \citet{de_simone_hot_2020} estimated an H$_2$ density of about $2\times 10^6$ cm$^{-3}$ in the inner $0\farcs24$. Using this gas density, the abundances of CH$_3$OH and SiO are  $\sim$10$^{-6}$ and $\sim$10$^{-8}$, respectively.

In summary, the observed emission arises from three zones: 
i) a very compact region ($<0\farcs15$, $<$50 au),  at $\pm 1.5$ \kms \ with respect to v$_{\rm sys}$, dominated by the hot corino emission and traced mainly by methanol;
ii) an intermediate region ($0\farcs15-0\farcs5$, 50--150 au ), between 2 and 6 \kms \ with respect to v$_{\rm sys}$, traced by both SiO and CH$_3$OH that shows an inverted velocity gradient, with respect to the large scale, whose origin is not clear;
iii) an extended region ($>0\farcs5$, $>150$ au), above $7$ \kms \ with respect to the v$_{\rm sys}$, traced by SiO and dominated by the jet/outflow.

\section{On the origin of the inverse velocity gradient}\label{sec:wide-outflow}

We now discuss the origin of the intermediate region in the inner $\sim0\farcs5$ ($\sim$150 au), where SiO and \meth \ show an observed inverse velocity gradient with respect to the large scale, and their observed spatial and spectral shift. 
%In the following we investigate the origin of this intermediate zone, if it is due to the sublimation of the ice mantles like in the hot corino, or by the grain sputtering like in shocked regions.
At this stage, it is important to highlight the formation mechanisms of these two species: 
CH$_3$OH is formed on the icy grain mantles via CO hydrogenation \citep[e.g.,][]{watanabe_efficient_2002, rimola_combined_2014} and it can be released into gas phase by either ice mantle sublimation 
or by grain mantle sputtering in mild shocks  {in star forming regions} \citep[e.g.,][]{flower_methanol_2010}.  
On the contrary, SiO is formed via the sputtering/shattering of the grain cores by strong shocks that release Si into the gas phase where it quickly oxygenates. It is estimated that up to $\sim$10\% of Si or SiO can also be frozen into the grain mantles, and therefore be released by grain mantle sputtering in mild shocks as \meth \
\citep[e.g.,][]{caselli_grain-grain_1997, schilke_sio_1997, gusdorf_sio_2008a, gusdorf_sio_2008b, guillet_shocks_2011}. 
Being the two species aligned along the same Position Angle (PA), it can be reasonable to think that they both come from from the base of the wind launched from the protostellar disk. However, CH$_3$OH is more compact and could also probe the very inner part of the envelope where it is released by thermal evaporation.
% I move that up...! 
%We observe the overlap of the emission of three zones: 
%i) a very compact region ($<0\farcs1$) dominated by the hot corino emission and traced mainly by methanol;
%ii) an extended region ($>0\farcs5$) traced by SiO and dominated by the jet/outflow; 
%iii) an intermediate region ($0\farcs1-0\farcs5$) traced by both SiO and CH$_3$OH that shows the inverted velocity gradient.
%We may now wonder what is
%In the following we investigate the origin of this intermediate zone, if it is due to the sublimation of the ice mantles like in the hot corino, or by the grain sputtering like in shocked regions. 
We investigate the following three possibilities:

\paragraph{\textit{Rotating disk/inner envelope:}} 
%At first glance, t
The blue- and red- shifted emission of CH$_3$OH might indicate rotation in a disk-like structure and/or in the inner envelope, where the ice sublimation dominates the chemistry.
{However, 
i) the PA ($\sim -30^\circ$) of the velocity gradient differs of $\sim30^\circ$ from the perpendicular direction of the jet/outflow direction at large scale (above 300 au), as expected for disk emission. Indeed, the jet/outflow direction is estimated to have a PA of $25^\circ-30^\circ$ \citep[see figure \ref{fig:sio_largescale}, and][Chahine et al. submitted]{choi_variability_2005, santangelo_jet_2015, de_simone_seeds_2020};}
% {the PA ($\sim -30^\circ)$) of the velocity gradient is *30 gradi diverso alla perperdicolare del *not  perpendicular to the jet/outflow direction (PA between 25-30 degrees; Figure \ref{fig:sio_largescale}, choi, santangelo, de simone, chahine et al. submitted); }
ii) { unfortunately, the present spatial resolution ($\sim$ 50 au) is not enough to assess whether the signature of a rotating/infalling envelope is present in CH$_3$OH and/or SiO (see Figure \ref{fig:pvs_sio+ch3oh} in appendix)};
iii) CH$_3$OH and SiO are aligned on the same PA, and both trace the inverted velocity gradient, but only 
\meth \ is a known disk tracer. 
%In case of a rotating disk/inner envelope, the species would be sublimated from the icy grain mantles. 

Indeed, while methanol emission can be explained by ice sublimation, for SiO is not so straightforward.
At the methanol emission site (extending out to 90 au, $0\farcs3$, and peaking at 45 au, $0\farcs15$, from the protostar) the dust temperature is 65--90 K , comparable to the CH$_3$OH sublimation one (60--130 K) (see computations in Appendix \ref{sec:sublim_O_sputt}). 
On the other hand,  SiO is usually associated to jet activity, i.e., where the Si is released by grain sputtering in shocks or entrained in a wind originated inside the dust sublimation radius  { \citep[e.g., ][]{cabrit_2007,gusdorf_sio_2008b, hirota_disk-driven_2017, tabone_constraining_2020,  lopez-vazquez_multiple_2024}}. %, and not observed in hot corinos
%\footnote{Please note that SiO has always being considered a typical shock tracer and never considered a hot corino tracer. Therefore no dedicated observations have been performed in this respect.}. 
{Even if there is 10\% of SiO trapped in the ice mantles \citep[e.g.,][]{gusdorf_sio_2008a,gusdorf_sio_2008b,guillet_shocks_2011}, it is not clear if this would be released with the ice once sublimated due to SiO reactivity with water or due to other processes (see details in Appendix \ref{sec:sublim_O_sputt}). 
%While single-dish observations do not discard the possibility that SiO is released by ice mantle sublimation \citep{ceccarelli_structure_2000}, interferometric maps of SiO, in hot corinos with jets, show a velocity gradient along the jet PA \citep[e.g.,][]{podio_calypso_2021, van_der_wiel_alma-pils_2019}. }
%puoi riportare che nelle hot corino sources quando risolta l’emissione SiO mostra sempre un velocity gradient lungo il PA del jet (quindi perpendicolare al disco) suggerendo che il bulk dell’emissione e’ dovuta al jet (i.e. sputtering negli shocks, and/or molecular formation in a wind originating from inside the dust sublimation radius, Tabone et al. 2020)
Alternatively, the two species could be released by grain mantle sputtering at the infalling envelope-disk interface, where slow shocks ($<$ 5-10 \kms) occur \citep[e.g.,][]{sakai_chemical_2014}. However, it is unclear if this slow shocks can sputter enough SiO, and if so, the emission would follow the disk kinematics, for which we do not have strong evidence (see above). 

In summary, the observed methanol emission could be the result of the ice mantle sublimation in the inner rotating envelope/disk, while a fast shock ($>$10 \kms) needs to occur to sputter the dust grains so to explain the observed SiO abundance. 
%Additionally, the overall geometry does not support this scenario.
%We may want to stress that the situation is more complicated and difficult to estimate with the current knowledge. Further investigation from both observation and chemical side are needed.   

\paragraph{\textit{Jet/Outflow precessing and/or on the plane of the sky:}} 
Another possibility is that \meth \ and SiO have a non-thermal origin linked to shocks along the jet.
The inclination of the IRAS 4A2 jet is yet highly uncertain \citet[45$^\circ$-88$^\circ$][]{yildiz_apex_2012,choi_variability_2005,marvel_time_2008,koumpia_evolutionary_2016}.
If it is close to the plane of the sky  
% (inclination lower than opening angle)
one expects to observe blue- and red-shifted emission in both lobes and along the whole jet length, not only in the inner 300 au \citep[see e.g.,][]{cabrit_co_1990, podio_calypso_2021}. 
On the other hand, if the jet precesses and crosses the plane of the sky it may create a knot with opposite velocity with respect to previous ejection events. However, if this was the case we would expect to see an alternation of blue- and red- shifted emission with a periodic pattern and also at larger scales. Moreover, this scenario has been excluded by \citet{chuang_alma_2021} based on a 3D precession model. Indeed, this could explain the S-shape structure of the IRAS 4A outflows traced in SO at large scale, but not the inversion of the velocity gradient at small scales.

\paragraph{\textit{Wide-angle wind:}} 
Figure \ref{fig:sketch-jet-outflow} shows a simplistic sketch that summarize our interpretation of the observed spatial-kinematical distribution.  
The sketch shows the base of a wide and inclined disk wind that crosses the plane of the sky and shows an inversion in velocity at 300 au with respect to the large-scale jet/outflow. 
{The disk wind is plausibly magnetically driven and it is expected to reach the maximum radius at the Alfv\'en surface and to re-collimate beyond that \citep{tabone_constraining_2020}. The prototypical disk wind in HH212 has been detected at heights below  $\sim$200 au \citep{franck_jets_2014,tabone_alma_2017, lee_first_2021, nazari_alma_2024}, consistent with our findings.   }
%The material crossing the plane of the sky will have high inclination needed to support the sputtering scenario. Indeed, 
Given the covered velocity ranges between 2 and 6 \kms, the inclination of the gas crossing the plane of the sky has to be $>55^\circ$ with respect to the line of sight. This is needed to have a reasonable flow velocity ($>$ 10 \kms) to sputter grain mantles and cores so to release both CH$_3$OH and SiO (see Appendix \ref{sec:sublim_O_sputt}). 

Among the three scenarios, we favour the last one as it explains the inverse velocity gradient, the spatial alignment between SiO and \meth, and also the observed SiO and \meth \ spectra (where the emission close to v$_{\rm sys}$ is hot corino dominated and beyond 7 \kms \ is jet/outflow dominated). 
For the first time, we detect and identify a disk wind candidate toward the IRAS 4A Class 0 protostar traced in SiO and \meth. 
This increases the number of disk wind tracers, complementing the typical SO \citep[e.g.,][]{tabone_alma_2017, codella_water_2018} and supporting the recently discovered \meth \ \citep{nazari_alma_2024}, and gives important constraints (velocity, inclination, and opening angles) for disk wind theories.
 {It is important to notice that the emission of SiO can also be the result of the interaction of the jet with the disk wind giving rise to an unresolved shocked shell \citep[see e.g.,][]{lee_magnetocentrifugal_2022}, or a disk wind rotation \citep[e.g.,][]{lee_magnetocentrifugal_2022}. In this way, SiO can be considered as an indirect tracer. However, to explain the observed low velocities, the jet/wind inclination needs to be quite high (above 80$^{\circ}$ with respect to the line of sight; see Appendix \ref{sec:sublim_O_sputt}). Anyways, observations down to at least 10 au would be needed to test this scenario. }

%These two species were not considered as disk wind tracers, and they will then now complement the previously identified disk wind tracers, as SO \citep[e.g.,][]{tabone_alma_2017, codella_water_2018}.

%The contribution of a disk cannot be completely excluded, but if present, it is unresolved in our observation, therefore with a size below 50 au. 

\begin{figure}
    \centering
    %\begin{minipage}[]{0.55\textwidth}
%    \includegraphics[width=1\columnwidth]{Sketch_jet-outflow_half.png}
    \includegraphics[width=\columnwidth]{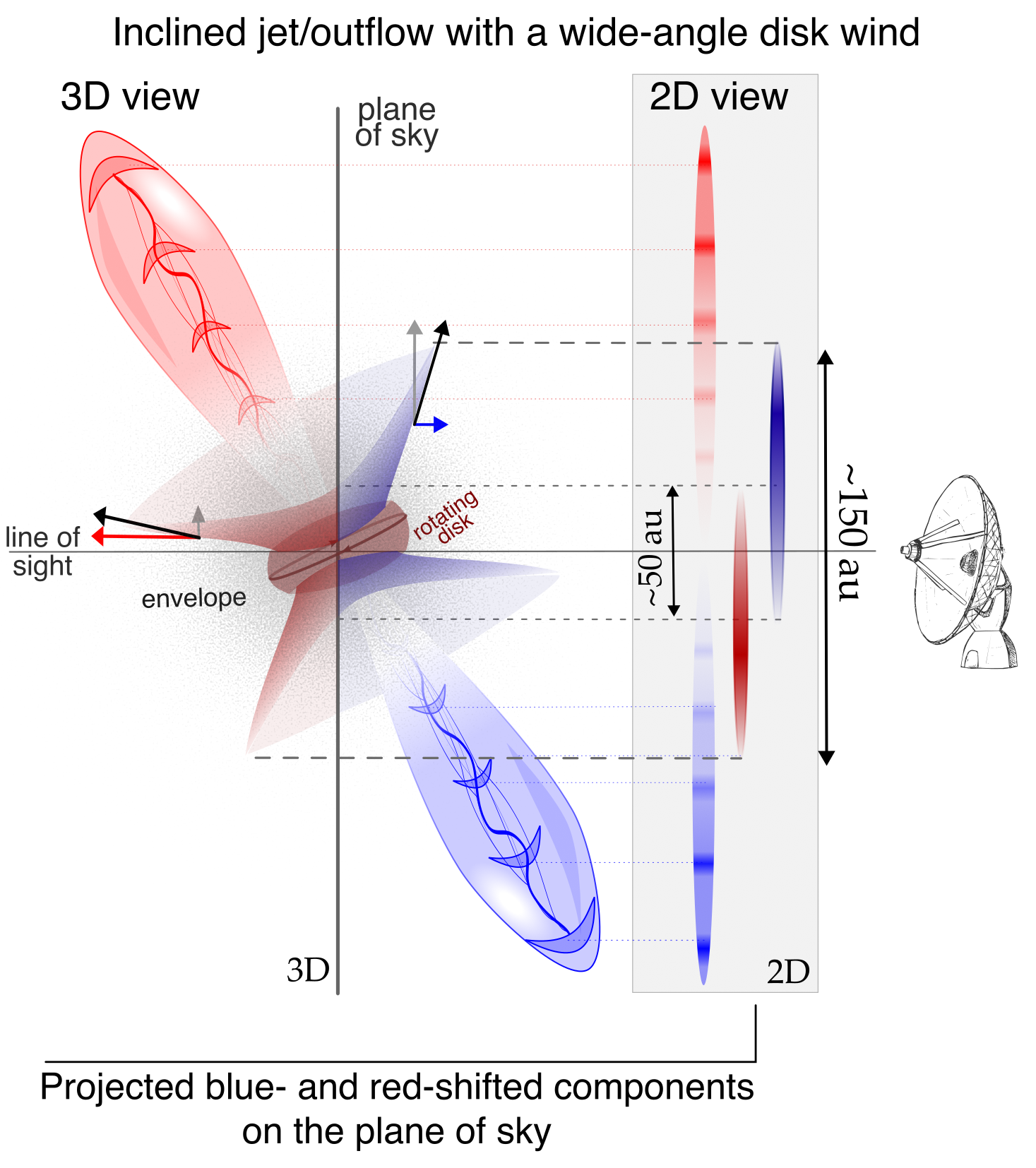}
%    \end{minipage}\hfill
%    \begin{minipage}[]{0.25\textwidth}
    \caption{Simplistic sketch (not in scale) of the proposed scenario: 
    %\textit{Left:} 
    The large scale jet is quite inclined (60$^{\circ}$ in the example), and it has a wide outflowing shocking gas that crosses the plane of the sky.
    %\textit{Right:} The large scale jet is almost on the plane of the sky crossing it partially. {Remove arrows rotation in disk!}
    The red-shifted blue-shifted emission are represented by red and blue colors respectively. The gray-shadowed box highlight the view of the projected emission on the plane of the sky that we actually observe.   }
    \label{fig:sketch-jet-outflow}
 %   \end{minipage}
\vspace{-0.5cm}
\end{figure}

\section{Conclusions}
This Letter presents FAUST observations with 50 au resolution of IRAS 4A2 as mapped in SiO and CH$_3$OH. 
The high spatial and spectral resolution observation of FAUST have been crucial to study the interplay between the disk, the envelope and the outflowing gas. The unique combination of kinematics and chemistry allowed us to disentangle the emission at very small scales (50 au) using the chemical formation pathway of each species to retrieve the physical mechanisms at play.  
The emission below 300 au arises from three zones: 
%i) a very compact region dominated by the hot corino emission and traced only by methanol; ii) an extended region traced by SiO and dominated by the jet/outflow;iii) an intermediate region traced by both SiO and CH3OH that shows an inverted velocity gradient. Indeed, while the large scale outflows are blue-shifted in the south and red-shifted in the north and covers a broad range of velocities (more than 30 \kms \ with respect to the v$_{\rm sys}$), the inner 300 au are characterised by a red-shifted emission in the south and blue-shifted emission in the north covering a smaller range of velocity (max 6 \kms \ with respect to the v$_{\rm sys})$.
i) a very compact region ($<0\farcs1$, 50 au),  at $\pm 1.5$ \kms \ with respect to v$_{\rm sys}$, dominated by the ice mantle sublimation zone and traced by methanol;
ii) an intermediate region ($0\farcs1-0\farcs5$, 50-150 au), between 2 and 6 \kms \ with respect to v$_{\rm sys}$, traced by both SiO and CH$_3$OH that shows an inverted velocity gradient with respect to the large scale, with unknown origin;
iii) an extended region ($>0\farcs5$, 150 au), above 7 \kms \ with respect to v$_{\rm sys}$, traced by SiO and dominated by the outflow.
We propose that \meth \ and SiO in the intermediate region probe the base of a rotating wide-angle disk wind. The material accelerated in the wind crosses the plane of the sky, giving rise to the observed inverted velocity gradient, and sputter the grains releasing both species. 
For the first time we observe the IRAS 4A disk wind  traced by, SiO and \meth, new tracers beyond the known SO. 
Finally, we emphasise that observations down to 10 au will be essential to reveal the presence and the possible contribution of the unresolved rotating disk. % and complete the overall picture of the system.  

\paragraph{\textit{Acknowledgements}: }
We warmly acknowledge Dr. J. Enrique Romero for the fruitful discussion on the binding energies and the Si, SiO insights, and Dr. F. Bacciotti on disk winds.
This project has received funding from the Marie Sklodowska-Curie for the project “Astro-Chemical Origins” (ACO), grant agreement No 811312, and within the European Union’s Horizon 2020 research and innovation program from the European Research Council (ERC) for the projects “The Dawn of Organic Chemistry” (DOC), grant agreement No 741002.
Part of the data deconvolution and analysis was performed using the GRICAD infrastructure (\url{https://gricad.univ-grenoble-alpes.fr}).
ClCo, LP, and GS acknowledge the PRIN-MUR 2020  BEYOND-2p (Astrochemistry beyond the second period elements, Prot. 2020AFB3FX), the PRIN MUR 2022 FOSSILS (Chemical origins: linking the fossil composition of the Solar System with the chemistry of protoplanetary disks, Prot. 2022JC2Y93), the project ASI-Astrobiologia 2023 MIGLIORA (Modeling Chemical Complexity, F83C23000800005), the INAF-GO 2023 fundings PROTO-SKA (Exploiting ALMA data to study planet forming disks: preparing the advent of SKA, C13C23000770005), the INAF Mini-Grant 2022 “Chemical Origins” (PI: L. Podio), and INAF-Minigrant 2023 TRIESTE (“TRacing the chemIcal hEritage of our originS: from proTostars to planEts”; PI: G. Sabatini).
D.J. is supported by NRC Canada and by an NSERC Discovery Grant.
L.L. acknowledges the support of DGAPA PAPIIT grants IN108324 and IN112820 and CONACyT-CF grant 263356”.
M.B. acknowledges support from the European Research Council (ERC) Advanced Grant MOPPEX 833460.
N.C acknoeledges funding from the European Research Council (ERC) under the European Union Horizon Europe research and innovation program (grant agreement No. 101042275, project Stellar-MADE).
This paper makes use of the following ALMA data: ADS/JAO.ALMA\#2018.1.01205.L (PI: S. Yamamoto). ALMA is a partnership of the ESO (representing its member states), the NSF (USA) and NINS (Japan), together with the NRC (Canada) and the NSC and ASIAA (Taiwan), in cooperation with the Republic of Chile. The Joint ALMA Observatory is operated by the  ESO, the AUI/NRAO, and the NAOJ.

%TC:ignore
\appendix
\counterwithin{figure}{section}

\section{Overview of FAUST SiO at large scale}
Figure \ref{fig:sio_largescale} shows the large scale SiO emission of the IRAS 4A system integrated in different velocity ranges. 
This provides a general overview of the whole range of velocities overall involved, for the sake of completeness.  Please note that the SiO emission is actually going beyond the spectral window bandwidth, therefore beyond the upper velocity range value reported in the figure. It is important to note for the present work that the northern part is red-shifted and the southern part is blue-shifted. 
The study of the large scale structure is out of the scope of this letter and presented in Chahine et al. (submitted).

\begin{figure}
    \centering
    \includegraphics[width=0.95\columnwidth]{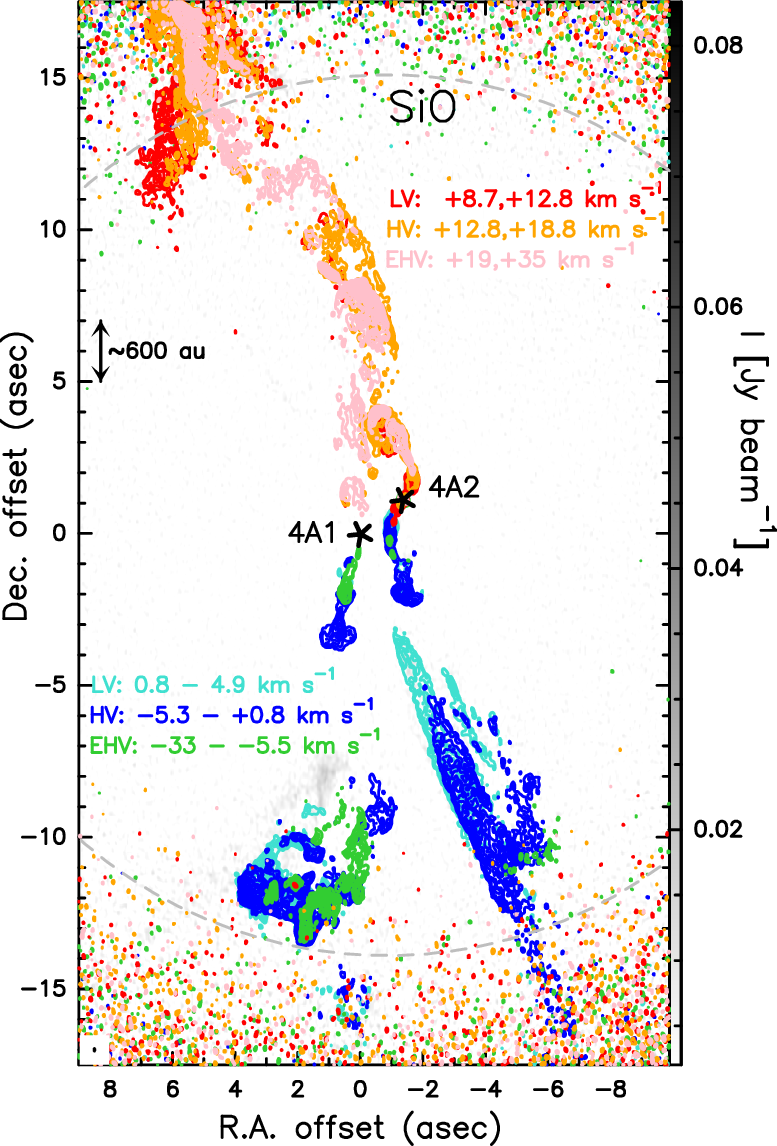}
    \caption{Large scale view of the ALMA FAUST SiO emission. Each velocity range corresponds to a specific color shown in contours starting at 4$\sigma$ with steps of 2$\sigma$, where $\sigma$= 3, 3.5, 6 mJy beam$^{-1}$ km s$^{-1}$ for the LV (low velocity: red and cyan), HV (high velocity: orange and blue), EHV (extremely high velocity: pink and green) range, respectively. In background, in grey scale, a range around the v$_{\rm sys}$ is shown (4.9--8.5 km s$^{-1}$). The synthesised beam is depicted in the lower left corner, and the primary beam is shown as a dashed grey circle.  }
    \label{fig:sio_largescale}
\end{figure}

\section{Channel maps}
Figure \ref{fig:sio_chanmaps} and \ref{fig:ch3oh-chanmaps} show the channel maps of SiO and CH$_3$OH, respectively. The channels are binned to 1 km s$^{-1}$, for the sake of clarity, and are zoomed in the inner 300 au of the IRAS 4A2 protostar. 
The inverted velocity gradient region is present only in the $(-6,-2)$ and $(2,6)$ \kms \ ranges.

\begin{figure*}
    \centering
    \includegraphics[width=\textwidth]{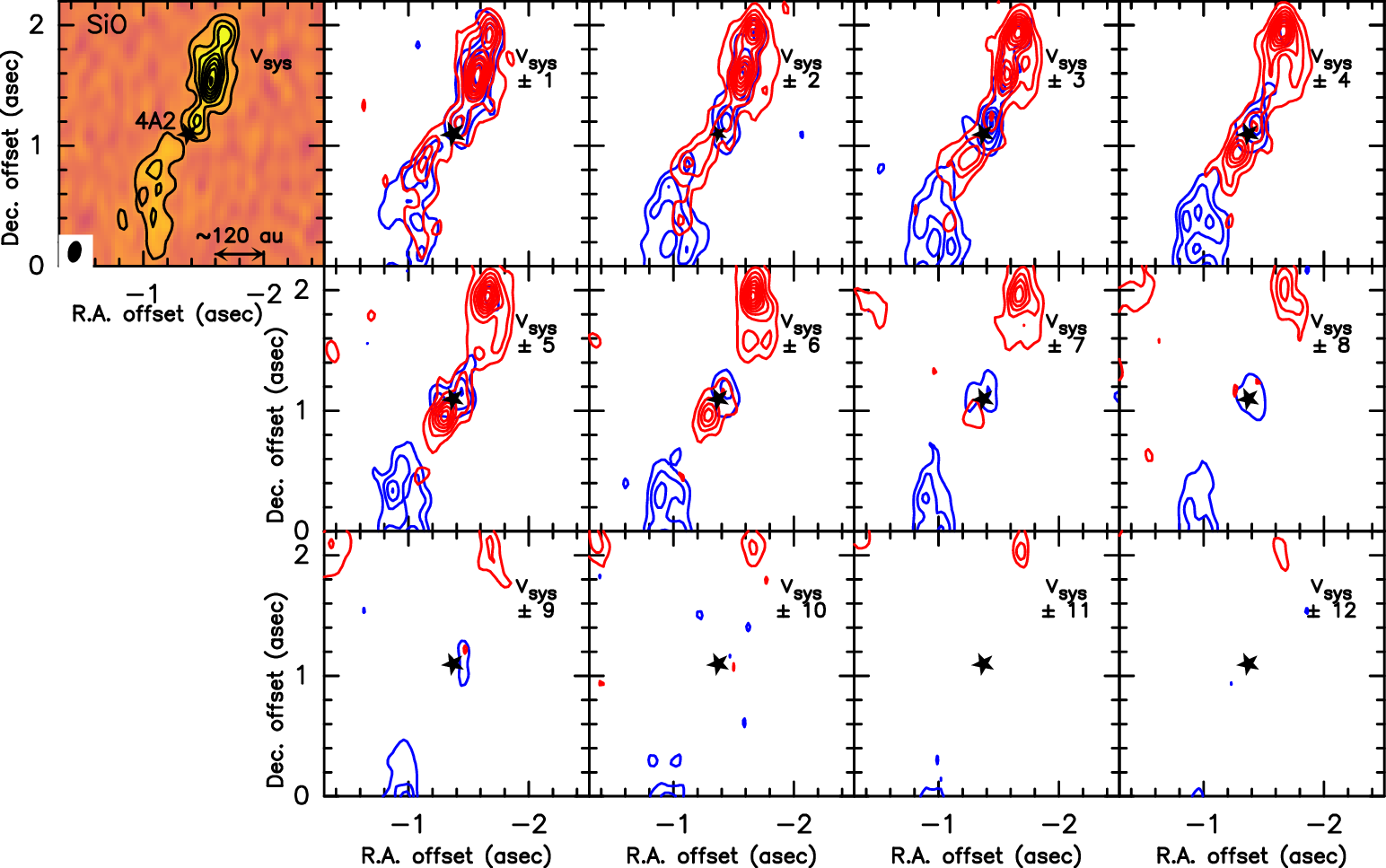}
    \caption{ALMA FAUST SiO velocity channel maps. Each panel corresponds to a 1 km s$^{-1}$ increment from the systemic velocity \citep[$\rm v_{sys}=6.8$ km s$^{-1}$;][]{choi_high-resolution_2001}.  The black star marks the position of the IRAS 4A2 protostar. The SiO beam is reported in the lower left corner of the $\rm v_{sys}$ panel (shown in color scale). First contours and steps are 3$\sigma$ ($\sigma=1$ mJy beam$^{-1}$).}
    \label{fig:sio_chanmaps}
\end{figure*}
\begin{figure*}
    \centering
    \includegraphics[width=\textwidth]{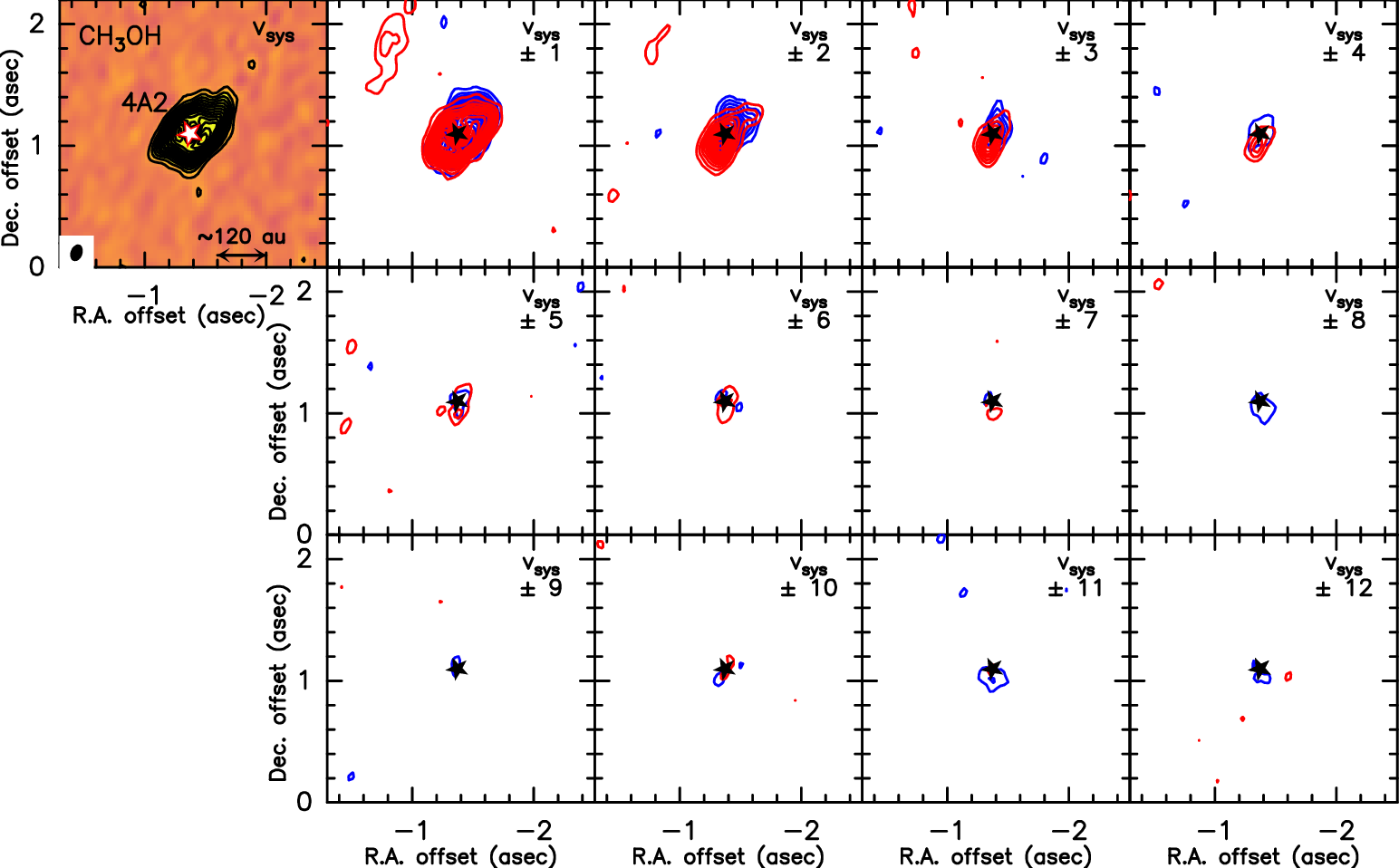}
    \caption{ALMA FAUST CH$_3$OH velocity channel maps. Each panel corresponds to a 1 km s$^{-1}$ increment from the systemic velocity \citep[$\rm v_{sys}=6.8$ km s$^{-1}$;][]{choi_high-resolution_2001}.  The black star marks the position of the IRAS 4A2 protostar. The CH$_3$OH beam is reported in the lower left corner of the $\rm v_{sys}$ panel (shown in color scale). First contours and steps are 3$\sigma$ ($\sigma=1.1$ mJy beam$^{-1}$).}
    \label{fig:ch3oh-chanmaps}
\end{figure*}

\section{SiO and CH$_3$OH: sputtering or sublimation?}\label{sec:sublim_O_sputt}
%Sio created mainly by sputtering because  {blalblabla}... From sublimation could be only a 10\%, could explain the intensity? maybe check abundances using C18O?
%As explained in the main text, SiO is thought to be formed mainly by fast oxidation of Si released into the gas by sputtering/shattering of the dust grains. On the other hand, methanol, grain-mantle species, is thought to be released in the gas by both ice mantle sublimation and/or sputtering/shattering. 

\paragraph{Sublimation:}
In order to have the desorption of a species from the ice mantle, the dust needs to reach the species sublimation temperature. 
%This process is the thermal desorption of a species bound to a substrate and it can be described approximately by the Polanyi-Wigner equation (Kolasinski 2002). 
Considering species that are not strongly associated to each other (e.g, CH$_3$OH adsorbed on a surface made mostly by water or CO) the first order solution for the desorption rate ($\rm k_{des}$) is: 
\begin{equation}\label{eq:desorp_rate}
    \rm 
    k_{des}=\nu_{des}e^{-BE[K]/T} \ ,
\end{equation}
where BE is the binding energy (namely, the strength of a species to remain attached to the surface) in Kelvin, and
$\nu_{des}$ is the pre-exponential factor which depends on the species and the surface  \citep{minissale_thermal_2022, ferrero_acetaldehyde_2022}.
Even if the BE is given most of the time as a single value, recent experiments and theoretical computations have shown that for each species there can be a distribution of BEs \citep{ferrero_binding_2020, bovolenta_high_2020,minissale_thermal_2022, Tinacci_theoretical_2022, tinacci_theoretical_2023}. This is because BE can change based on the site and orientation of the species with respect to the molecules on the substrate.

To retrieve the sublimation temperature from the desorption rate (i.e., the BE value) we solve Equation \ref{eq:desorp_rate} taking the half-life time $\rm t_{1/2}$ as the characteristic time of the desorption, in first approximation, as \citep{ceccarelli_organic_2023}:
\begin{equation}
  \rm t_{1/2}=\frac{\ln(2)}{k_{des}}= \frac{\ln(2)}{\nu_{des}e^{-BE[K]/T}} \ .
\end{equation}
Therefore,
\begin{equation}
\rm    T_{sub}= \frac{BE[K]}{\ln(t_{1/2})+\ln(\nu_{des}/\ln(2))} \ .
\end{equation}

Regarding methanol, theoretical predictions by \citet{ferrero_binding_2020} indicate a BE range of (3770--8618) K, while the pre-exponential factor $\rm \nu_{des}$ is $3.2\times 10^{17}$ s$^{-1}$ \citep{minissale_thermal_2022}. 
The BE given by \citet{minissale_thermal_2022} is 6621 K, consistent with the range found by \citet{ferrero_binding_2020}.
Using the CH$_3$OH BE range and $\nu_{des}$ stated above, and assuming that the grain mantles have been warmed up for a timescale of $\sim10^4$ yr\ \footnote{Please note that assuming 1 Myr timescale, $T_{sub}$ could be lower by about 10 K.}, $\rm T_{sub}=$ $60-130$ K. 

We can now estimate the dust temperature at a certain distance from the protostar.
When the dust is optically thin, the dust temperature profile heated by a central source with  L$_\star$ luminosity, can be approximated by the following equation \citep{ceccarelli_structure_2000}:
\begin{equation}
\rm T_{dust}[K]= 75 K \ \Big(\frac{L_\star}{27 L_\odot}\Big)^{1/4}\Big(\frac{r}{150au}\Big)^{-1/2} \ .
\end{equation}
    
The bolometric luminosity (L$_{\rm bol}$ ) of IRAS 4A1 and IRAS 4A2 is not measured singularly due to the lack of angular resolution observation at infrared wavelengths. The total bolometric luminosity of the IRAS 4A1+4A2 system is estimated to be 9 L$_\odot$ \citep{kristensen_water_2012, karska_water_2013}. 
The blue- and red- shifted emission peak of methanol are separated by about $0\farcs15$ (about 45 au) and their emission extend up to 0$\farcs26$ away (about 80 au). If we consider for 4A2 conservatively a bolometric luminosity of 4.5 L$_\odot$ (half of the total luminosity), the temperature at 80 au from the protostars is about 65 K, while at 45 au is about 90 K. 
%In summary, the temperature at the distance from the protostar where we observe methanol could be enough to sublimate CH$_3$OH from the grain mantles.  

Regarding SiO, it could be possible to investigate if sublimation can play a role, given that it is thought that about 10\% of SiO can be already formed and locked into the ice mantles of the dust grains \citep[see e.g., ][]{gusdorf_sio_2008a, gusdorf_sio_2008b, guillet_shocks_2011}. %Therefore, the sublimation of the ice mantles for the high dust temperature could lead to the release into the gas phase of this percentage of SiO.
The binding energy of SiO is reported to be about 3500 K \citep{hasegawa_new_1993}, that translate into a sublimation temperature of about 50 K. Therefore, in theory, at $0\farcs2$ (60 au) away from the protostar (at SiO emission peak), the dust temperature could be enough to sublimate SiO. 
If we consider the computed \meth/SiO column density ratio (see Table \ref{tab:spectral properties}), we find that the abundances of SiO should be two order of magnitude lower than methanol. In IRAS 4A2, \citet{de_simone_hot_2020} estimated a gas density of about $2\times10^6$ cm$^{-3}$ in the inner $0\farcs24$. Using this gas density, the abundances should be about 10$^{-6}$ for CH$_3$OH and 10$^{-8}$ for SiO \citep[similar to the typical values in shocked regions; e.g.,][]{cabrit_2007,gusdorf_sio_2008b, taquet_seeds_2020}.
Considering a 10\% of elemental silicon frozen into SiO on the mantle, we could have an abundance of frozen SiO or Si on the icy mantle of about 10$^{-7}-10^{-8}$ \citep[see Figure 4 in][]{ceccarelli_evolution_2018,lodders_solar_2019}. 
Therefore, we could, in principle, explain the observed abundance. 
However, SiO and Si could be highly reactive with the water molecules in the ice preventing the sublimation of pure SiO and Si into the gas phase. New quantum chemical computations are ongoing to obtain 
i) updated values for the SiO binding energy taking into account orientation and ice morphology (Gelli et al. in prep.), and ii) evaluating how much reactive the SiO and Si are on interstellar ices (Enrique-Romero et al. private comm.). 
%However, we find difficult to think that pure SiO can easily and directly sublimate by the ice mantle (when T$_{\rm dust}$ is about 50 K), otherwise we would expect it to be ubiquitous in all protostars. 

\paragraph{Sputtering:}
SiO is the best known and most studied shock tracers since more than two decades. Several shock models have predicted that the SiO abundance is highly enhanced in high velocity shocks (above 25 \kms). Indeed at these velocities the silicon is liberated from the refractory cores of the grains thanks to sputtering and/or shattering, and goes into the gas phase where it quickly oxidizes \citep[e.g.,][]{flower_grain_1994, caselli_grain-grain_1997, schilke_sio_1997}. 
At lower velocities, some species (including Si, directly SiO, and also CH$_3$OH) that were previously frozen on to the grain mantles can still be released into the gas phase by the sputtering of the icy mantles itself \citep{jimenez-serra_parametrization_2008, gusdorf_sio_2008b, guillet_shocks_2011, lesaffre_low_2013, nguyen-luong_low_2013}. The above models predict that, to liberate Si or SiO from the frozen mantles, a minimum velocity of about 10 \kms \ is needed\footnote{Please note that these values critically depend on the assumed frozen-SiO sputtering threshold energy \citep[see discussion in][]{de_simone_train_2022, flower_grain_1994}}. 

From our observations, the small scale inverse velocity component has a velocity range between 2 and 6 \kms \ (see Figure \ref{fig:cont+sio+ch3oh}, \ref{fig:sio_chanmaps}, \ref{fig:ch3oh-chanmaps} ). This represent the line-of-sight velocity component. To retrieve the radial velocity, that would be the responsible for the sputtering, we need information on the jet/outflow inclination. 
The IRAS 4A jet inclination is not very well known yet. 
Indeed, due to the large extent of the outflows/jets and the high line-of-sight velocities \citet{yildiz_apex_2012} suggested an inclination of $45-60^\circ$ with respect to the line of sight. On the other hand, \citet{choi_variability_2005}, \citet{marvel_time_2008}, and \citet{koumpia_evolutionary_2016} suggested larger inclination of about $79^\circ$, $88^\circ$ and 70$^\circ$, respectively.
However, we can give a constrain on the inclination of the emission at this scale (not necessarily of the jet, see discussion in Section \ref{sec:wide-outflow}) based on the minimum velocity needed to have sputtering, so to explain the presence of both SiO and CH$_3$OH. 
Considering that the observed line-of-sight velocity is $2,6$ \kms \ with respect to the v$_{\rm sys}$, and the velocity required for sputtering a reasonable amount of SiO and CH$_3$OH (see above), we need an inclination with respect to the line of sight above 55$^{\circ}$ for a radial velocity of 10 \kms, and above 80$^\circ$ for 35 \kms.

\section{PV Diagrams}

\begin{figure}
    \centering
    \includegraphics[width=0.97\columnwidth]{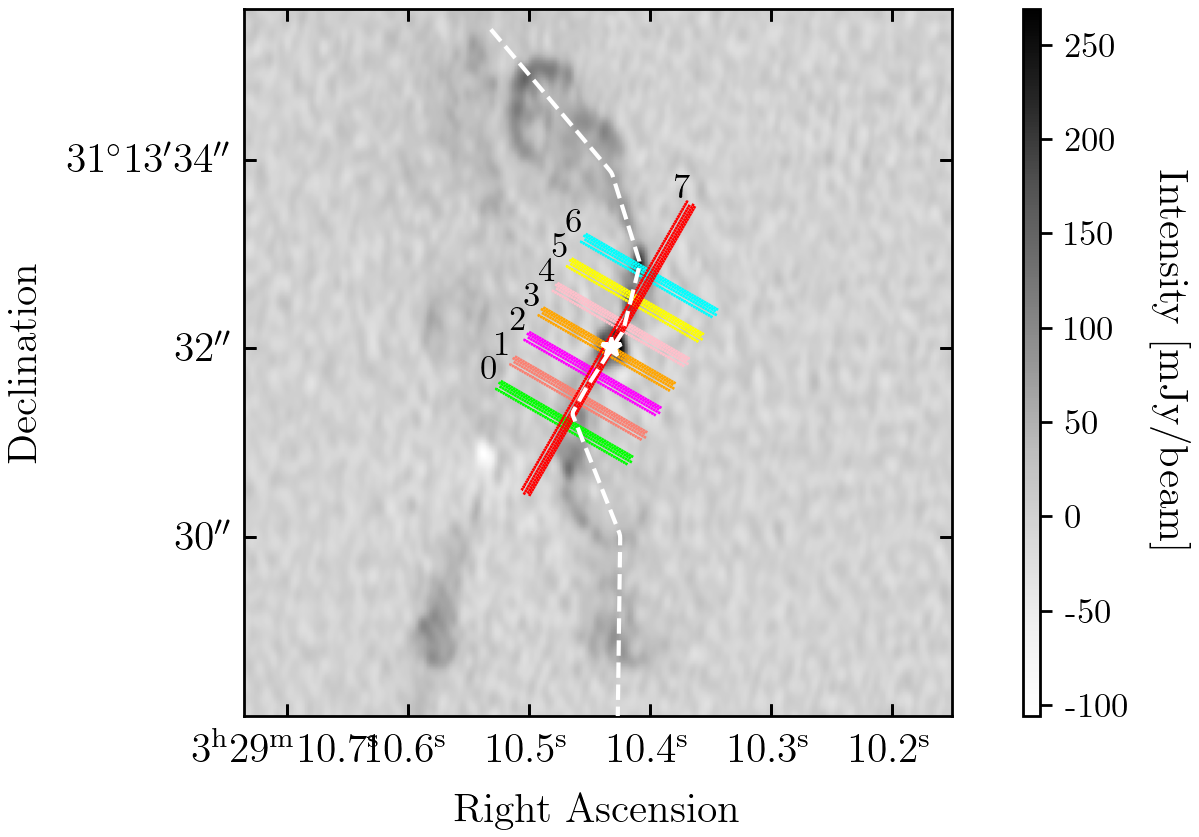}
    \caption{Regions (colored) where PV diagrams have been extracted. 
    %The red slice is of $3\farcs6$ length passing through the direction identified by the red- and blue-shifted emission peak of methanol in the inner $0\farcs5$  (PA$\sim$30$^{\circ}$). The other colored lines are perpendicular to the flow axis (i.e. slices perpendicular to the methanol and SiO PA), of $1\farcs6$ length and separate by about $0\farcs3$. The white line marks the path along the whole SiO structure below $4''$ from 4A2. All regions are $0\farcs1$ wide. 
    For reference, the moment 0 of SiO integrated over the whole spectral window is in grey scale. IRAS 4A2 is marked by the white star.}
    \label{fig:regions_pv}
\end{figure}
\begin{figure*}
    \centering
    \includegraphics[width=0.95\columnwidth]{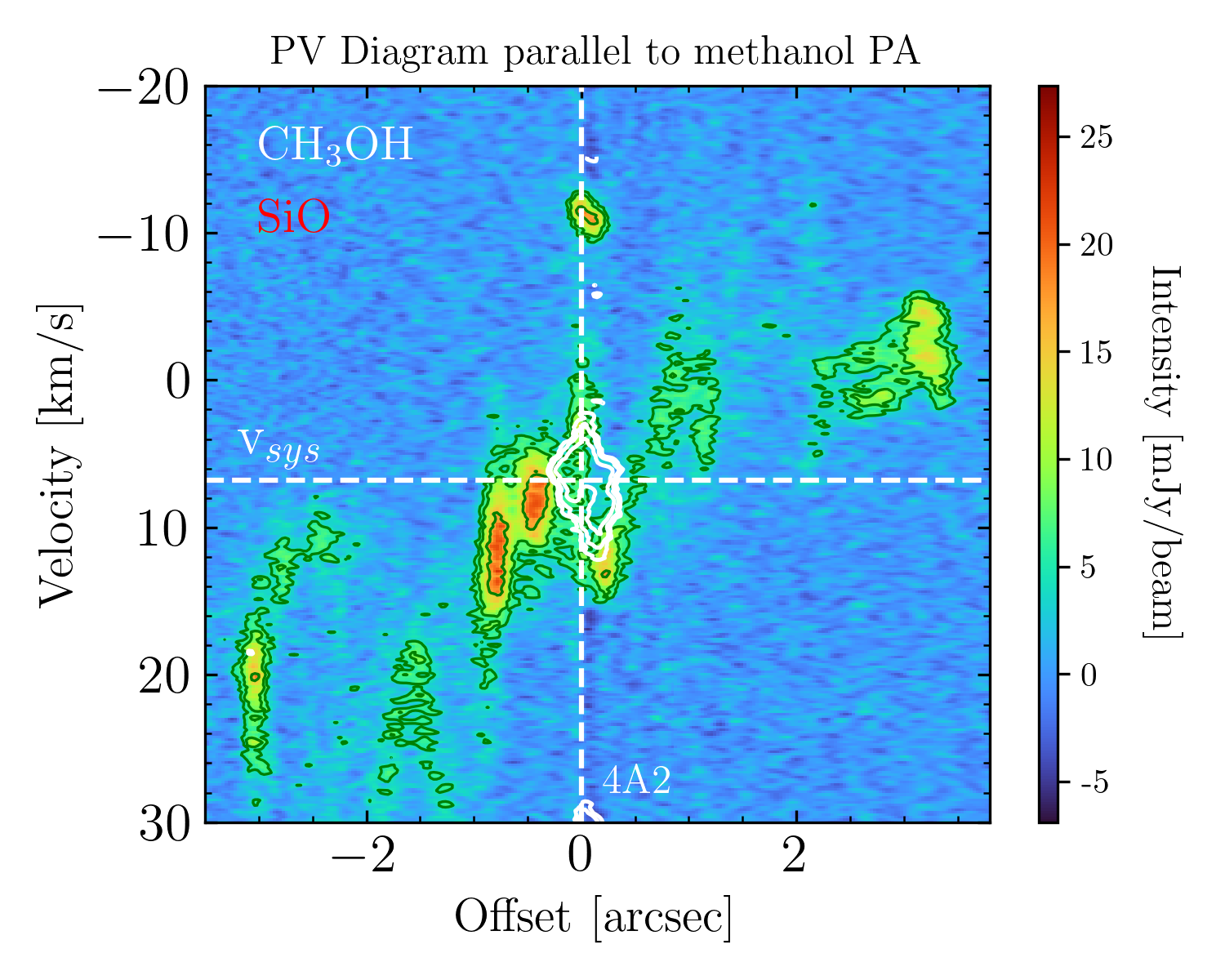}
       \includegraphics[width=0.95\columnwidth]{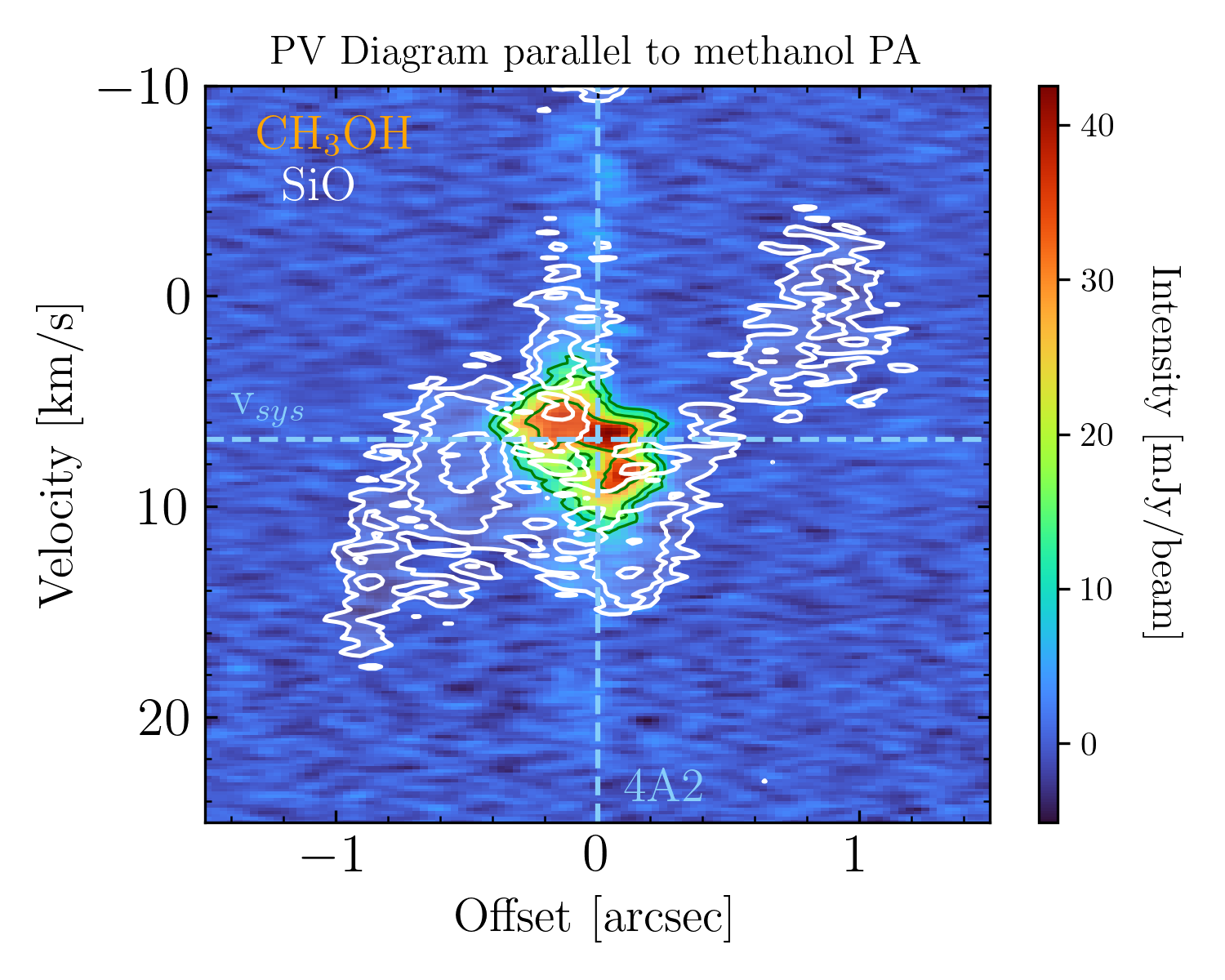}
    \caption{
     {\textit{Left:}  PV Diagram extracted over the path of $\sim7''$ length that follows the SiO structures (white in Figure \ref{fig:regions_pv}). The SiO emission is shown in color scale.  Green and white contours at (3,5,10) $\sigma$ are for \meth \ ($\sigma\sim2$ mJy beam$^{-1}$) and SiO ($\sigma_m\sim$1.6 mJy beam$^{-1}$), respectively. 
    \textit{Right} PV Diagram extracted over the region of 3$''$ length passing through methanol and SiO PA in the inner $0\farcs5$ (shown in red in the left panel). The methanol emission is shown in color scale with green contours (at 3,5,10 $\sigma_m$ with $\sigma_m\sim$3 mJy beam$^{-1}$). The SiO emission is shown in white contours at (3,5,10, 20) $\sigma_s$ with $\sigma_s\sim$1.5  mJy beam$^{-1}$. 
    In both panels, the vertical and horizontal dashed lines represent the v$_{\rm sys}$ \citep[$\sim$6.8 \kms;][]{choi_high-resolution_2001}, and the 0$''$ offset (on the 4A2 protostar), respectively. 
    }
    }
    \label{fig:pvs_sio+ch3oh}
\end{figure*}

\begin{figure*}
    \centering
    \includegraphics[width=0.95\textwidth]{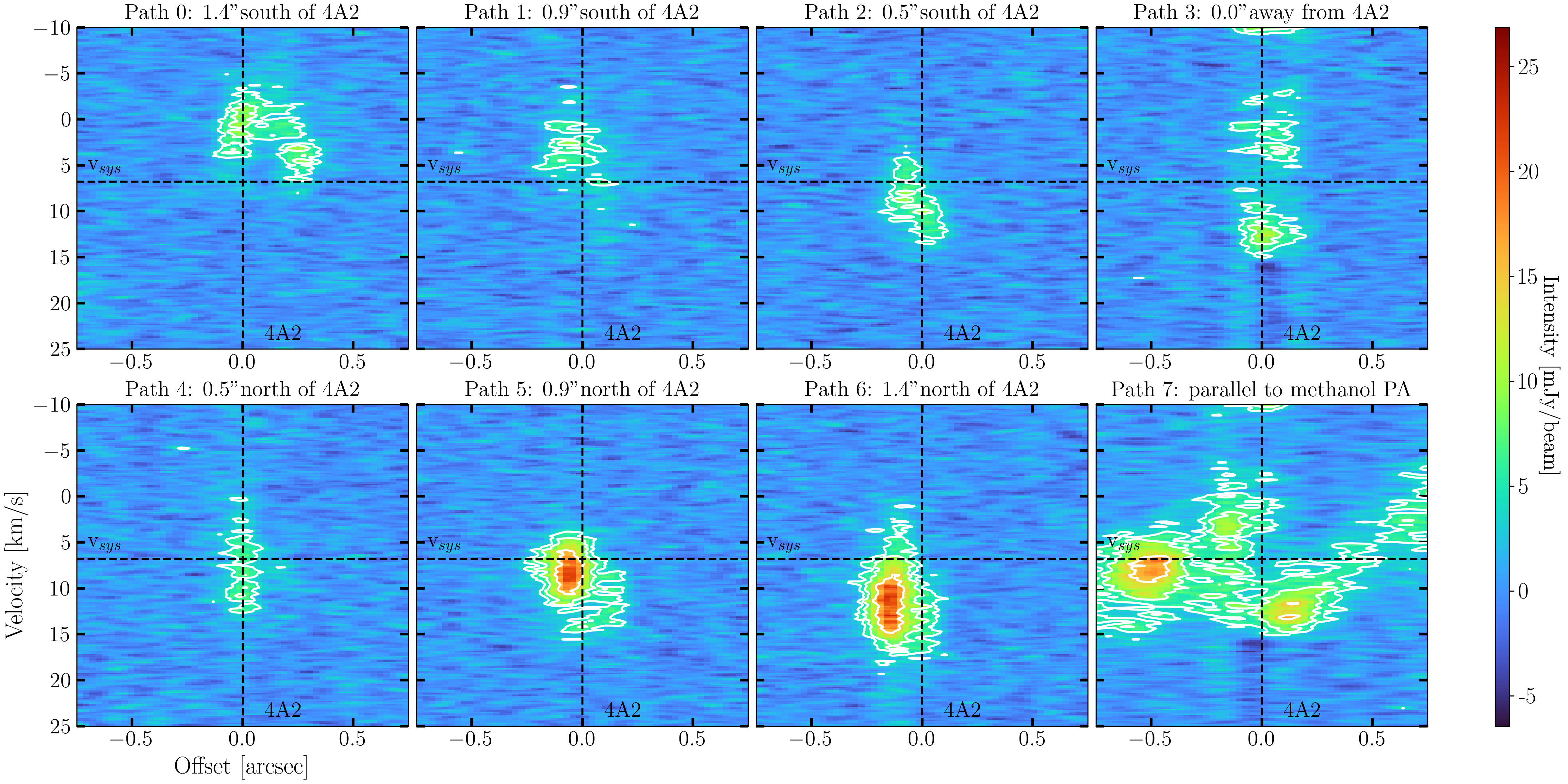}
    \caption{ {SiO PV diagrams extracted at the region shown in Figure \ref{fig:regions_pv} and opportunely labelled at the top of each plot. The contours are at (3,5,10, 20) $\sigma_s$ with $\sigma_s\sim$1.5  mJy beam$^{-1}$. The vertical and horizontal dashed lines represent the v$_{\rm sys}$ \citep[$\sim$6.8 \kms;][]{choi_high-resolution_2001}, and the 0$''$ offset (on the 4A2 protostar), respectively.  }}
    \label{fig:sio_pv_diagrams}
\end{figure*}

 {We selected eight regions $0\farcs1$ wide (shown in Figure \ref{fig:regions_pv}) where to extract the PV diagrams: one of $3\farcs6$ length passing through the direction identified by the red- and blue-shifted emission peak of methanol in the inner $0\farcs5$  (PA$\sim$30$^{\circ}$) and the other seven are perpendicular to that, each separated by $0\farcs3$, and of $1\farcs6$ length. The PV diagrams extracted from these regions for SiO are shown in Figure \ref{fig:sio_pv_diagrams}, while the one extracted from the parallel slice and the path following the SiO structures (shown in red and white, respectively, in Figure \ref{fig:regions_pv}) for both \meth \ and SiO are shown in Figure \ref{fig:pvs_sio+ch3oh}). 
It is possible to notice, as also visible in the channel maps (Figure \ref{fig:sio_chanmaps}), that the SiO PV diagram on the path $0\farcs$9 and $0\farcs$6 south of IRAS 4A2 are blue-shifted while $0\farcs$3 south a red-shifted component appears together with the blue-shifted ones. In the north paths the situation is similar with red and blue-shifted inverted. The emission is confined around $\pm 0\farcs3$ of the selected axis, and does not show any peculiar structure. Only the emission along the path $0\farcs$9 south of 4A2, that correspond to the cavity seen in the southern outflow (see Figure \ref{fig:cont+sio+ch3oh}), is resolved and shows higher velocities with respect to the small scale structures (below $0\farcs5$). 
In Figure \ref{fig:pvs_sio+ch3oh}, the emission along the slice parallel to \meth \ and SiO PA (red in Figure \ref{fig:regions_pv}) of SiO shows a more evident structure. 
It is noticeable the inversion of velocity gradient, with red emission in the south and blue emission in the north, below $0\farcs5$ from 4A2. The knot at $-0\farcs5$ can be interpreted as the first knot associated to the outflow. Indeed, the left panel of the figure shows the PV diagram extracted from a region that follows the SiO structure in the moment 0 (white line in Figure \ref{fig:regions_pv}). Here two knots are at $\sim0\farcs5$ and $\sim1''$ with 8.5 \kms \ and 12 \kms \ respectively and indicate knots in the outflow. 
Comparing SiO and \meth \ in Figure \ref{fig:pvs_sio+ch3oh} right), we can see that \meth \ is elongated following the SiO emission with inverted velocity gradient. However, we are limited by spatial resolution to reveal any other kinematical structure. 
}

%TC:endignore
\bibliographystyle{aa}
\bibliography{DeSimone_IRAS4A_diskwind}{}

\end{document}